\documentclass[12pt]{article}
\usepackage{amsmath,latexsym}
\usepackage{graphicx}
\begin{document}
\title{\textbf{Wormholes supported by polytropic phantom energy}}
\author{
\small Mubasher Jamil\footnote{mjamil@camp.edu.pk}, Peter K. F.
Kuhfittig\footnote{kuhfitti@msoe.edu}, \small Farook
Rahaman$^1$\footnote{$^1$farook\_rahaman@yahoo.com}  and
 Sk. A Rakib$^\ddag$ \\ \\
 $^\ast$\small Center for Advanced Mathematics and Physics,
National University of Sciences and Technology,\\ \small E\&ME
campus, Peshawar Road, Rawalpindi - 46000, Pakistan
\\
$^\dag$\small Department of Mathematics, Milwaukee School of
Engineering,
Milwaukee, Wisconsin 53202-3109, USA\\
$^\ddag$\small Department of Mathematics, Jadavpur University,
Kolkata - 700032,
 India\\
\small \\
 }\maketitle
\date{}
\begin{abstract}\noindent
It is generally agreed that the acceleration of the
Universe can best be explained by the presence of dark or phantom
energy.  The equation of state of the latter shows that the null
energy condition is violated.  Such a violation is the primary
ingredient for sustaining traversable wormholes.  This paper
discusses wormholes supported by a more general form called
polytropic phantom energy.  Its equation of state results in
significant generalizations of the phantom-energy and, in
some cases, the generalized Chaplygin-gas wormhole models, both
of which continue to receive considerable attention from
researchers.  Several specific solutions are explored, namely,
a constant redshift function, a particular choice of the shape
function, and an isotropic-pressure model with various shape
functions.  Some of the wormhole spacetimes are asymptotically
flat, but most are not.
\end{abstract}
\newpage

\section{Introduction}

A typical stationary spherically symmetric wormhole is a two-mouthed
tunnel (also called a tube, throat, or handle) in a 
multiply-connected spacetime joining two remote asymptotically flat 
regions of the same spacetime or two different spacetimes altogether.  
The wormhole concept has been extended to time-dependent rotating
wormholes \cite{teo}, cylindrical wormholes \cite{ger}, wormholes
with a cosmological constant \cite{lemos1}, plane symmetric
thin-shell wormholes \cite{lemos}, and charged wormholes \cite{kim}.
(See \cite{lobo} for a review). Historically, the concept of a
wormhole was suggested by Flamm in 1916 \cite{flamm} (shortly 
after the publication of the Einstein field equations) by means 
of the now standard embedding diagram.  A similar construction 
was later attempted by  Einstein and Rosen \cite{albert},
the so-called Einstein-Rosen bridge.  It was subsequently shown 
that the bridge, also called a ``Schwarzschild wormhole," is 
really a black hole \cite{wheeler}. The
concept of ``traversable wormhole" was suggested by Morris and
Thorne \cite{morris} with the idea of using a wormhole for
interstellar travel or even time travel. Wormholes could also be
enlarged through a mechanism similar to cosmological inflation
\cite{roman}. Wormholes appear to lead to a violation of the Hawking
chronology protection conjecture and give faster-than-light
scenarios to observers outside the wormhole \cite{hawking},
assuming, of course, that the above-mentioned geometry in the
Einstein field equations yields the matter distribution that is
intrinsically inhomogeneous and ``exotic." The matter is termed
exotic since it violates the standard energy conditions (null and
weak) that are generally obeyed by classical matter. In other words,
its energy density takes on negative values in some suitable
reference system, as does the pressure. It turns out that injecting
this exotic matter into the wormhole's throat can significantly
widen the radius of the throat \cite{krasnikov}.

Recent interest has centered on matter that satisfies the equation
of state (EoS) $p=-\omega\rho$ with $\omega>0$.  There are several
possibilities: quintessence ($1/3<\omega<1$), phantom energy
$(\omega>1)$, and the case $\omega=1$, which is equivalent to the
presence of a cosmological constant.  Recent astrophysical
observations of supernovae of type Ia and cosmic microwave
background data suggest that the observable universe is pervaded by
a dynamic dark energy whose EoS parameter $\omega$ ranges from
sub-negative to super-negative values. Moreover, it comprises more
than seventy percent of the critical density required for a flat
universe \cite{perl}, suggesting that wormholes could form and 
stabilize in a dark-energy dominated universe
\cite{valerio}. In the opposite direction, if the exotic matter is
removed from the wormhole's throat, then the wormhole could decay
into a black hole \cite{peter}. It has been shown that the wormhole
region that requires exotic matter can be made arbitrary small by
introducing a special shape function \cite{kuhf}. Some time later,
attempts were made to devise wormhole solutions that do not require
a violation of energy conditions by modifying general relativity;
examples are Gauss-Bonnet gravity, Kalb Ramond background,
braneworld scenarios, and Brans-Dicke gravity theory \cite{luis}. In
the cosmological context, wormholes could also present a promising
solution to the famous horizon problem \cite{david}.

Wormholes are not purely mathematical entities but possess
astrophysical significance as well. In \cite{Kardashev}, it is
argued that wormholes can accrete ordinary matter that may convert
them into black holes with strong monopole magnetic fields.  In 
\cite{eduard}, it is shown that static Morris-Thorne wormholes 
can also cause gravitational lensing. Hence wormholes could be 
detected by their specific magnification curves, in particular, 
by observing the position of the peak of the Einstein's ring. 
In \cite{abd}, it is proposed that oscillations of bodies in 
the vicinity of a wormhole throat could give rise to a
peculiar observational phenomenon: signals from such sources
detected by an external observer will display a characteristic
periodicity in their spectra. All objects (stars and black holes)
other than wormholes irrecoverably absorb bodies falling onto them .
Periodic radial oscillations are a characteristic feature of
magnetized wormholes. In a separate study \cite{harko}, it is shown
that if a wormhole is surrounded by a thin accretion disk of matter,
then the potential well is deeper than the Schwarzschild potential,
and consequently more energy is radiated away from the wormhole.

In this paper we investigate static wormholes supported by
phantom energy satisfying a polytropic equation of state. On a
large cosmic scale, the phantom energy manifests itself as the
source causing a rapid accelerated expansion of the universe. It
can also lead to a cosmic singularity commonly called the ``big
rip" that is predicted to occur in a finite cosmic time, thereby
ripping apart every gravitationally bound structure, including
black holes \cite{jamil}. It thereby leads to the violation of
the laws of thermodynamics with the decrease in entropy and
violation of energy conservation \cite{odintsov}. Recently,
several authors have discussed traversable wormholes supported by
phantom energy  \cite{sushkov, fL05}.

This paper discusses wormholes supported by polytropic phantom
energy, resulting in significant generalizations of the
phantom-energy and, in some cases, the generalized
Chaplygin-gas models \cite{fL06}.  These generalizations could
have far-reaching cosmological implications, as future
observations may show that the generalized EoS provides a better 
fit, thereby increasing the probability of detecting
naturally occurring astronomical phenomena, including wormholes.
For example, it is shown in Ref. \cite{BP05} that the
generalized Chaplygin EoS not only points to the existence of
``Chaplygin dark stars," but expressions for their size and
expansion rates can be derived. The compatibility of such bodies
with the Chaplygin-gas cosmological background is also analyzed.
Similarly, in the framework of phantom-type matter in cosmology,
it is ascertained in Ref. \cite{CPRS09} that the theoretical
model, based on the Noether symmetry approach, is compatible with
presently available observational data.

\section{Modeling the system}\label{S:model}
Let us consider a Morris-Thorne wormhole, which is a static
spherically symmetric spacetime with line element
\begin{equation}
ds^2=-e^{2\Phi(r)}dt^2+\left(1-\frac{b(r)}{r}\right)^{-1}dr^2
  +r^2(d\theta^2+\sin^2\theta d\phi^2).
\end{equation}
This metric involves two arbitrary functions of the radial
coordinate $r$, $\Phi(r)$ and $b(r)$, called the \textit{redshift}
and the \textit{shape functions}, respectively. The former function
gives information about the gravitational redshift of an infalling
object, while the latter specifies the shape of the wormhole when
viewed, for example, in an embedding diagram. In the metric
coefficient $g_{rr}$, the minimum radius $r=r_0$ is termed the
\textit{throat} of the wormhole, where $b(r_0)=r_0$. The absence of
a horizon is necessary for the traversability of the wormhole. So to
get a reasonable wormhole geometry, some restrictions are imposed on
the metric coefficients $g_{tt}$ and $g_{rr}$: the redshift function
$\Phi$ must be finite for all values of $r$ to avoid an event
horizon. The shape function $b$ must satisfy the $flare$-$out$
condition at the throat: $b^\prime(r_0)<1$, in addition to
$b(r_0)=r_0$. Finally, the spacetime must be asymptotically flat at
large distances.

\emph{Remark 1:} Since a wormhole is assumed to join two different 
spacetimes, there actually exist two different redshift and shape 
functions, $\Phi_{\pm}$ and $b_{\pm}$, respectively.  A description 
of the wormhole now requires two coordinate patches, each covering 
the range $[r_0, +\infty)$, one in each universe.  The patches are 
joined at $r=r_0$.  For simplicity we assume that $\Phi_{+}=
\Phi_{-}$ and $b_{+}=b_{-}$, but this requirement is not essential 
to the definition of traversable wormhole.

Since we have now assumed the wormhole geometry, we need to solve
the Einstein field equations to specify the form of matter that will
support this wormhole and make it stable against the inward
gravitational force. The field equations are
\begin{eqnarray}
\frac{b^\prime}{r^2}&=&8\pi\rho,\\
2\left( 1-\frac{b}{r}
\right)\frac{\Phi^\prime}{r}-\frac{b}{r^3}&=&8\pi p_r,\\
\left( 1-\frac{b}{r} \right)\left[
\Phi^{\prime\prime}+(\Phi^\prime)^2+\frac{\Phi^\prime}{r}-\frac{b^\prime
r-b}{2r^2(r-b)}-\frac{b^\prime
r-b}{2r(r-b)}\Phi^\prime\right]&=&8\pi p_t.
\end{eqnarray}
Here $\rho$ is the energy density, while $p_r$ and $p_t$ are the
radial and transverse pressures, respectively, of the matter
content.  Observe that the three independent field equations
involve five parameters: $\rho$, $p_r$, $p_t$, $\Phi$ and $b$. To
solve the system exactly, we are going to make use of the following
ansatz, an equation of state describing \emph{polytropic phantom
energy}, which is a generalization of the earlier model:
\begin{equation}\label{E:polytropic}
p_r=-\omega\rho^{\gamma},\ \ \gamma=1+\frac{1}{n},\ \ n\neq0.
\end{equation}

Various specializations will be discussed in the sections below.

\section{A model with constant redshift function}\label{S:constant}

If $\Phi(r)\equiv c$, a constant, then we get from Eqs. (2), (3),
and (\ref{E:polytropic}),
\begin{equation}\label{E:constant}
   -\frac{b}{r^3}=-\frac{\omega\cdot 8\pi}{(8\pi)^{1+1/n}}
     \left(\frac{b'}{r^2}\right)^{1+1/n}.
\end{equation}
This equation can be solved by separation of variables:
\[
   b(r)=\left(\frac{1}{3}\frac{(8\pi)^{1/(n+1)}}
   {\omega^{n/(n+1)}}r^{3/(n+1)}+c\right)^{n+1},\quad n\ne -1.
\]
Observe that for very large $r$, $c$ becomes negligible and we
assume it to be zero. The result is
\[
   b(r)=\frac{1}{3^{n+1}}\frac{8\pi}{\omega^n}r^3.
\]
Since $b(r)/r$ does not go to zero, the spacetime is not
asymptotically flat.  We will return to this point later.

From the condition $b(r_0)=r_0$, we now get
\[
    b(r)=\left(\frac{1}{3}\frac{(8\pi)^{1/(n+1)}}
   {\omega^{n/(n+1)}}r^{3/(n+1)}
   +r_0^{1/(n+1)}
      -\frac{1}{3}\frac{(8\pi)^{1/(n+1)}}
        {\omega^{n/(n+1)}}r_0^{3/(n+1)}\right)^{n+1},
\]
while the flare-out condition $b'(r_0)<1$ yields
\begin{equation}\label{E:flare1}
    b'(r_0)=\frac{(8\pi)^{1/(n+1)}}
      {\omega^{n/(n+1)}}r_0^{2/(n+1)}<1.
\end{equation}
As a check against the phantom-energy case, let
$n\rightarrow\pm\infty$.  Then
\[
    b'(r_0)=\frac{1}{\omega}<1,
\]
and $\omega>1$, as expected.  So to be consistent with the
phantom-energy model, we assume that $\omega>1$ for $n>0$
and $n<-1$.  The remaining case, $-1<n<0$, leads to an
extension of the Chaplygin-gas model; here $\omega<1$,
as we will see.

Returning to the limiting case $n\rightarrow \pm\infty$,
Eq. (\ref{E:constant}) leads to
\begin{equation}\label{E:powerform}
    b(r)=r_0\left(\frac{r}{r_0}\right)^{1/\omega},
\end{equation}
which is Lobo's solution \cite{fL05}.  Observe that the
resulting spacetime is now asymptotically flat.

\subsection{$n>0$}\label{S:ng0}

In (\ref{E:flare1}), the parameter $n$, and hence $\gamma$, all
depend on $\omega$ and $r_0$ to meet the flare-out conditions. In
other words,
\begin{equation}\label{E:flare2}
    b'(r_0)=\left(\frac{8\pi r_0^2}
     {\omega^n}\right)^{1/(n+1)}<1
\end{equation}
implies that
\[
       n>\frac{\text{ln}(8\pi r_0^2)}{\text{ln}\,\,\omega}.
\]
Returning to Eq. (\ref{E:polytropic}), to meet the flare-out
conditions, we must have
\[
     \gamma<1+\frac{1}{N},\quad \text{where}\quad N=
      \frac{\text{ln}(8\pi r_0^2)}{\text{ln}\,\,\omega}.
\]
The implication is that $\gamma$ may not be much larger
than unity.  The resulting EoS is nevertheless an extension
of the phanton-energy model.

\subsection{$n<0$}\label{S:nl0}

Suppose we first consider the interval $-1<n<0$, which
includes the generalized Chaplygin-gas model $-1<n\le -1/2$.
From the flare-out condition we get
\[
   (8\pi r_0^2)^{1/(n+1)}<\omega^{n/(n+1)},
\]
which implies that
\begin{equation}\label{E:flare3}
   \omega<\frac{1}{(8\pi r_0^2)^{-1/n}}.
\end{equation}
Writing the EoS in the form
\[
      p_r=-\omega\frac{1}{\rho^{-1-1/n}}
\]
and noting that $-1-1/n>0$, let $a=-1-1/n$, so that $-1/n=a+1$.
Inequality (\ref{E:flare3})  then includes the condition
for generalized Chaplygin wormholes in Lobo's notation
\cite{fL06}:
\[
   \omega<\frac{1}{(8\pi r_0^2)^{a+1}}.
\]

As in Lobo's Chaplygin model \cite{fL06}, the condition
$b(r)<r$ implies that the wormhole cannot be arbitrarily large.
To keep $n+1$ positive, we assume now that $n>-1$.  Thus
\[
   b(r)=\left[\frac{1}{3}\frac{(8\pi)^{1/(n+1)}}
   {\omega^{n/(n+1)}}\left(r^{3/(n+1)}-
     r_0^{3/(n+1)}\right)+r_0^{1/(n+1)}
      \right]^{n+1}<r,
\]
and
\[
    \frac{1}{3}\frac{(8\pi)^{1/(n+1)}}
   {\omega^{n/(n+1)}}\left(r^{3/(n+1)}-
     r_0^{3/(n+1)}\right)<r^{1/(n+1)}
     -r_0^{1/(n+1)}.
\]
Factoring the left side and reducing, we get
\[
  \frac{1}{3}\frac{(8\pi)^{1/(n+1)}}
   {\omega^{n/(n+1)}}\left(r^{2/(n+1)}+
     r_0^{1/(n+1)}r^{1/(n+1)}
     +r_0^{2/(n+1)}\right)<1,
\]
which leads to the quadratic inequality
\[
     r^{2/(n+1)}+r_0^{1/(n+1)}r^{1/(n+1)}
     +r_0^{2/(n+1)}-\frac{3\omega^{n/(n+1)}}
         {(8\pi)^{1/(n+1)}}<0.
\]
The solution is
\[
   r^{1/(n+1)}<\frac{1}{2}\left(-r_0^{1/(n+1)}
    +\sqrt{r_0^{2/(n+1)}-4\left(r_0^{2/(n+1)}
      -\frac{3\omega^{n/(n+1)}}{(8\pi)^{1/(n+1)}}\right)}\right).
\]
To create dimensionless quantities, we divide by $r_0$:
\begin{equation}\label{E:Lobo1}
    \frac{r}{r_0}<\left[\frac{-1+\sqrt{-3+12\left(
    \frac{\omega^n}{8\pi r_0^2}\right)^{1/(n+1)}}}{2}\right]
       ^{n+1}.
\end{equation}
In view of conditions (\ref{E:flare2}) and (\ref{E:flare3}),
the dimensionless quantity
\begin{equation}\label{E:Lobo2}
   \left(\frac{\omega^n}{8\pi r_0^2}\right)^{1/(n+1)}
\end{equation}
is bigger than unity.  It follows that $r$ lies in the
range $r_0< r<\alpha r_0$ for some $\alpha>1$.  If $n=-1/2$,
condition (\ref{E:Lobo1}) is identical to Lobo's condition.
In summary, the EoS $p_r=-\omega\rho^{1+1/n}$ for
$-1<n<0$ leads to an extension of the generalized Chaplygin
wormhole with similar properties.

To study the condition $n<-1$, let $\gamma=1-\frac{1}{n}$
with $n>1$.  Then
\[
    \frac{r}{r_0}<\left[\frac{-1+\sqrt{-3+12\left(
    \frac{\omega^{-n}}{8\pi r_0^2}\right)^{-1/(n-1)}}}{2}\right]
       ^{-(n-1)}.
\]
which implies that $r/r_0<1$.  The case $0<\gamma<1$ is therefore
excluded.  So whenever the redshift function is constant,
either $\gamma>1$ or $\gamma \in (-1, 0)$.

As a final remark, since $\Phi'\equiv 0$, the expressions for
$\rho$, $p_r$, and $p_t$ are readily obtained from Eqs.
(2)-(4).


\section{A shape function in the form of a power}\label{S:power}

If $b(r)=b_0r^m$, $0<m<1$, we get from Eqs. (2)-(4)
\begin{equation}
\rho=\frac{b_0 m}{8\pi}r^{m-3}.
\end{equation}
\begin{equation}
p_r=-\omega\left( \frac{b_0m}{8\pi} \right)^{(n+1)/n}r^l,\ \
l=\frac{(n+1)(m-3)}{n}.
\end{equation}
\begin{equation}
\int d\Phi=\Phi=\Phi_0+\int
\frac{\frac{1}{2}b_0r^{m-2}-\frac{1}{2}B_0r^{l+1}}{1-b_0r^{m-1}}dr, \ \ B_0=8\pi
\omega\left( \frac{mb_0}{8\pi}  \right)^{(n+1)/n}.
\end{equation}

\subsection{The weak energy condition}\label{S:WEC}
Holding a wormhole open requires a violation of the weak energy
condition (WEC): $\rho+p_r<0$.  Recall that in the phantom-energy
case,
\[
   \rho+p_r=\rho-\omega\rho=\rho(1-\omega)<0,
\]
so that $\omega>1$.  The EoS $p_r=-\omega\rho^{1+1/n}$
yields a more complicated relationship:
\[
   \rho+p_r=\rho\left(1-\omega\rho^{1/n}\right)<0
      \quad \text{and} \quad 1-\omega\rho^{1/n}<0.
\]
From Eq. (2),
\begin{equation}\label{E:WEC1}
   1-\omega\left(\frac{1}{8\pi}\frac{b'(r)}{r^2}\right)
       ^{1/n}<0.
\end{equation}
So the WEC is violated whenever
\begin{equation}\label{E:WEC1a}
   \omega>\left(\frac{8\pi r^2}{b'(r)}\right)^{1/n}.
\end{equation}
At the throat, $b'(r_0)<1$, whence
\[
    \omega>(8\pi r_0^2)^{1/n}.
\]
As $n\rightarrow\pm \infty$, the phantom-energy case, this
reduces to $\omega>1$, as we saw earlier.

For the shape function $b(r)=b_0 r^m$, discussed in this section,
the requirement $b(r_0)=r_0$ leads to $b(r)=r_0^{1-m}r^m$.  Since
$b'(r_0)=m<1$, the flare-out condition is met.

From inequality (\ref{E:WEC1a}), we get
\begin{equation}\label{E:WEC2}
\omega>\left[\frac{8\pi r_0^2}{m}\left(\frac{r}{r_0}\right)
^{3-m}\right]^{1/n}
\end{equation}
As before, as long as $\omega$ meets the required condition,
the WEC is violated.

In Eq. (\ref{E:WEC2}), whenever $n>0$ a suitable model may
require $n$ to be quite large.  This will result in a value for
$\gamma$ close to unity and an EoS close to that of the
phantom-energy model.  An alternative approach allowing a smaller
$n$ (with the same shape function) will be discussed in Sec.
\ref{S:poweragain}, where the junction to an external
spacetime is considered.

\subsection{The redshift function}\label{S:redshift}

We saw that for the shape function $b(r)=r_0^{1-m}r^m$,
$0<m<1$, the flare-out condition is met, while
$b(r)/r\rightarrow 0$.  The special case $m=1/2$,
discussed by Lobo \cite{fL06}, leads to an
exact solution for $\Phi(r)$, but avoids an event horizon
only by assigning a specific value to $\omega$.  Since our
model contains the additional parameters $n$ and $m$, we
cannot obtain an exact solution, but, as shown below, an
event horizon can be avoided just the same.

From Eq. (3) we write
\[
   2\left(1-\frac{r_0^{1-m}r^m}{r}\right)\frac{\Phi'}{r}
   -\frac{r_0^{1-m}r^m}{r^3}=8\pi\left[-\omega
    \left(\frac{1}{8\pi}\frac{mr_0^{1-m}r^{m-1}}{r^2}\right)
       ^{(n+1)/n}\right].
\]
Solving for $\Phi'(r)$ and simplifying,
\begin{equation}\label{E:Phiprime1}
   \Phi'=\frac{r^{(3-m)/n}
   -\frac{\omega}{(8\pi)^{1/n}}m^{(n+1)/n}
        r_0^{(1-m)/n}}{r_0^{m-1}2rr^{(3-m)/n}
     \left(r^{1-m}-r_0^{1-m}\right)}.
\end{equation}
To eliminate an event horizon, we introduce the following
requirement in Eq. (\ref{E:Phiprime1}): let
\[
    \frac{\omega}{(8\pi)^{1/n}}m^{(n+1)/n}r_0^{(1-m)/n}
       =r_0^{(3-m)/n},
\]
which translates into the following condition on the radius of
the throat:
\begin{equation}\label{E:throat}
    r_0=\left(\frac{\omega}{(8\pi)^{1/n}}m^{(n+1)/n}\right)
       ^{n/2}.
\end{equation}
Eq. (\ref{E:Phiprime1}) can now be rewritten:
\begin{equation}\label{E:Phiprime2}
    \Phi'=\frac{1}{2r_0^{m-1}r^{(3-m+n)/n}}
     \times\frac{r^{(3-m)/n}-r_0^{(3-m)/n}}
        {r^{1-m}-r_0^{1-m}}.
\end{equation}
So by L'Hospital's rule,
\[
  \lim_{r \to r_0+}\Phi'(r)=\Phi'(r_0)=\frac{3-m}{2n(1-m)r_0}.
\]
It follows that $\Phi'(r)$ and hence $\Phi(r)$ are well
behaved at the throat.  Observe that in Eq. (\ref{E:throat}),
if we let $m=1/2$ and $n=-1/2$, and solve for $\omega$, we get
\begin{equation}
   \omega=\frac{1}{128\pi^2r_0^4}.
\end{equation}
This is Lobo's condition, based on the EoS for the Chaplygin
wormhole.  This restriction on the EoS is replaced by
Eq. (\ref{E:throat}).

Returning to Eq. (\ref{E:Phiprime2}), it is interesting to note
that if $n\rightarrow \pm\infty$, the phanton-energy case, then
$\Phi'\equiv 0$, which takes us back to Sec. \ref{S:constant}, Eq.
(\ref{E:powerform}).


\section{Models with isotropic pressure}\label{S:isotropic}
The conservation of the stress-energy tensor,
$T^{\hat{\mu}\hat{\nu}}_{\phantom{\mu\mu};\,\hat{\nu}}=0$,
gives
\[
    p'_r=\frac{2}{r}(p_t-p_r)-(\rho+p_r)\Phi'.
\]
Under the assumption of isotropic pressure, $p_r=p_t=p$, we obtain
\begin{equation}\label{E:iso1}
\rho=\frac{b^\prime}{8\pi r^2}.
\end{equation}
\begin{equation}\label{E:iso2}
p=-\omega\left(\frac{b^\prime}{8\pi r^2}\right)^{1+1/n}.
\end{equation}
\begin{equation}\label{E:iso3}
\Phi=\Phi_0-(n+1)\text{ln}\left|1-\omega\left(\frac{b^\prime}{8\pi
r^2} \right)^{1/n}\right|.
\end{equation}

In the remainder of this section we will examine this model more
closely using various shape functions.  Since we are  going to
concentrate primarily generalizing the phantom-energy model, we
will assume that $n>0$ or $n<-1$ with $\omega>1$.

\subsection{$b(r)=b_0r^m$, $0<m<1$}\label{S:poweragain}

In this section we return to the shape function
$b(r)=r_0^{1-m}r^m$.  From Eqs. (\ref{E:WEC2}) and
(\ref{E:iso3}),
\begin{equation}\label{E:redshift}
   \Phi(r)=\Phi_0-(n+1)\text{ln}\left|1-\omega\left[\frac{m}
    {8\pi r_0^2}\left(\frac{r_0}{r}\right)^{3-m}\right]
     ^{1/n}\right|;
\end{equation}
$\rho$ and $p$ can be obtained similarly.  We already saw that
whenever $\omega$ meets the conditions
discussed in Sec. \ref{S:WEC}, the WEC is violated.  Furthermore, for any
fixed $n$, the second term on the right-hand side of Eq.
(\ref{E:redshift}) goes to zero.  So $\Phi_0=0$.  Since $b(r)/r$
also goes to zero, the spacetime is asymptotically flat.

As noted earlier, for $n>0$, $n$ may have to be quite large for
$\omega>1$ to meet the required condition.  An alternative
allowing a smaller $n$ is based on condition (\ref{E:WEC2}):
\begin{equation}\label{E:WEC3}
  \omega>\left[\frac{8\pi r_0^2}{m}\left(\frac{r}{r_0}\right)
^{3-m}\right]^{1/n}
\end{equation}
Suppose $\omega$ is determined empirically.  At the throat
$r=r_0$, we have
\[
   \omega>\left(\frac{8\pi r_0^2}{m}\right)^{1/n}.
\]
To satisfy this inequality, we must let
\[
  n>n_1=\frac{\text{ln}\frac{8\pi r_0^2}{m}}
   {\text{ln}\,\omega},\quad \omega>1.
\]
However, this $n_1$ cannot satisfy inequality (\ref{E:WEC3}) for
$r>r_0$.  So the wormhole cannot be arbitrarily large, that is,
$r$ must be confined to some interval $[r_0, r_1]$.  The
condition on $n$ now becomes
\begin{equation}\label{E:N}
    n>N=\frac{\text{ln}\left[\frac{8\pi r_0^2}{m}
        \left(\frac{r_1}{r_0}\right)^{3-m}\right]}
     {\text{ln}\,\omega}
\end{equation}
and $\gamma<1+\frac{1}{N}$.  Observe that in view 
of Eq.(\ref{E:redshift}), inequality (\ref{E:N}) is exactly 
the condition needed for $\Phi(r)$ to be defined on $[r_0, r_1]$.

Since the resulting spacetime is no longer asymptotically flat, it
will have to be cut off at some $r=a$ and joined to an external
Schwarzschild spacetime.  In that case, $\Phi_0$ in Eq.
(\ref{E:redshift}) has to be determined from the junction
conditions. According to Ref. \cite{LLde}, a junction surface is a
thin shell.  Being infinitely thin in the radial direction, the
radial pressure is zero. The surface energy density is also zero,
but the tangential pressure is not.  (See Ref. \cite{LLde} for
details.) According to Ref. \cite{morris}, if a wormhole is to be
used for traveling, then the space station should be far enough away
from the throat for the spacetime to be nearly flat.  In other
words, $b(a)/a\ll 1$ for some $r=a$.  If $r_0=10$\,m and $m=1/2$,
then the choice $b(a)/a=0.001$ yields $a= 10000000$\,m, which is
also a convenient distance for the cut-off.

For $r>a$, the Schwarzschild line element becomes
\[
    ds^2=-\left(1-\frac{b(a)}{r}\right)dt^2+\frac{dr^2}
     {1-b(a)/r}+r^2(d\theta^2+\text{sin}^2\theta\,d\phi^2),
\]
where $b(a)/a\approx 0.001$.  From
\[
   \Phi(a)=\Phi_0-(n+1)\text{ln}\left|1-\omega\left[\frac{1}
     {1600\pi}\left(\frac{10}{a}\right)^{5/2}\right]^{1/n}\right|
        =\frac{1}{2}\text{ln}\left(1-\frac{b(a)}{a}\right),
\]
and, given $\omega$, we can now determine $\Phi_0$.  At $r=a$,
\[
  |\Phi(a)|\approx 0.0005\ll 1,
\]
also recommended in Ref. \cite{morris}.

If $n<-1$, the inequality (\ref{E:WEC3}) is trivially
satisfied and $\Phi(r)$ is defined.  For the resulting
$\gamma$, we get $0<\gamma<1$.  (The results in this section
can be extended to the interval $-1<n<0$, provided that
$\omega$ is sufficiently small, as, for example, in Eq.
(\ref{E:flare3}).)

\emph{Remark 2:} We have assumed for physical reasons that
$\omega>1$ in the phantom-energy model.  Mathematically,
this assumption is not necessary, however: if $n<-1$ and
$\omega<1$, then inequality (\ref{E:WEC3}) is satisfied
whenever $|n|<|N|$.

\emph{Remark 3:} The remaining subsections contain brief
numerical calculations that allow a comparison of the
various shape functions.  In particular, the distance
to the cut-off determines the required value of $n$.
The throat radii are obtained from $b(r_0)=r_0$, also for
comparison.  (All capitalized letters represent
constants.)

\subsection{$b(r)=A\text{tan}^{-1}(Cr)$}
Once again we start by listing $\rho$, $p$, and $\Phi$:
\begin{equation}
\rho= \frac{AC}{8\pi r^2(1+C^2r^2)}.
\end{equation}
\begin{equation}
p=-\omega\left( \frac{AC}{8\pi
r^2(1+C^2r^2)}\right)^{1+1/n}.
\end{equation}
\begin{equation}
\Phi=\Phi_0-(n+1)\text{ln}\left|1-\omega \left(
\frac{AC}{8\pi r^2(1+C^2r^2)} \right)^{1/n}\right|.
\end{equation}
The use of the shape function $b(r)=A\text{tan}^{-1}(Cr)$ offers
one important advantage over $b(r)=b_0r^m$: since
$b(r)/r\rightarrow 0$ more rapidly, the junction to an external
Schwarzschild spacetime can take place at a relatively small
value of $r$, thereby producing a smaller wormhole structure.
Consider, for example,
\[
   b(r)=3.5 \text{tan}^{-1}(0.55r).
\]
The condition $b(r_0)=r_0$ yields $r_0\approx 4.0$\, m for the
radius of the throat.  Also, $b'(r_0)<1$.  As expected, the
choice $b(a)/a=0.001$ results in a considerably smaller $a=
5500$\,m.

For the WEC discussed in the previous section,
\begin{equation}\label{E:WEC4}
   \omega\left(\frac{1.925}{8\pi r^2[1+(0.55)^2r^2]}\right)
       ^{1/n}>1.
\end{equation}
To satisfy this condition for $n>0$ in the interval $[r_0,a]$,
we find that
\[
   n>\frac{35.8}{\text{ln}\,\omega},\quad \omega>1,
\]
This value for $n$ is relatively small; so the resulting value
for $\gamma$ is relatively large, thereby producing a significant
generalization of the phantom-energy model.

As in the previous section, if $n<-1$, the inequality
(\ref{E:WEC4}) is satisfied for any empirically
determined $\omega$.  So once again, $0<\gamma<1$.

\subsection{$b(r)=r_0[1+\beta^2\left(1-\frac{r_0}{r}
\right)]$,\quad$\beta^2<1$}
 \begin{equation}
\rho=\frac{\beta^2r_0^2}{8\pi r^4}.
\end{equation}
\begin{equation}
p=-\omega\left( \frac{\beta^2r_0^2}{8\pi r^4} \right)^{1+1/n}.
\end{equation}
\begin{equation}
\Phi=\Phi_0-(n+1)\text{ln}\left| 1-\omega\left( \frac{\beta^2r_0^2}{8\pi
r^4} \right) ^{1/n}\right|.
\end{equation}

Here the results are somewhat similar to those using $b(r)=
A\text{tan}^{-1}(Cr)$: if $\beta=1/2$ and $r_0=10$\,m, then
$b(r)/r=0.001$ produces $a=12500$\,m for the cut-off.  The
condition
\[
    \omega\left(\frac{\beta^2r_0^2}{8\pi r^4}\right)
       ^{1/n}>1
\]
yields
\[
    n>\frac{37.7}{\text{ln}\,\omega},
\]
which is also similar.  As before, for any given $\omega$, the
value of $\Phi_0$ in the redshift function
\[
   \Phi=\Phi_0-(n+1)\text{ln}\left|1-\omega\left(
    \frac{\beta^2r_0^2}{8\pi r^4}\right)^{1/n}
        \right|
\]
can now be determined.

For the remaining shape functions the results are once again similar.

\subsection{$b(r)=D(1-\frac{A}{r})\left(1-\frac{B}{r}\right)$}
\begin{equation}
\rho=\frac{-2ABD+(A+B)Dr}{8\pi r^5}.
\end{equation}
\begin{equation}
p=-\omega\left(\frac{-2ABD+(A+B)Dr}{8\pi
r^5}\right)^{1+1/n}.
\end{equation}
\begin{equation}
\Phi=\Phi_0-(n+1)\text{ln}\left| 1-\omega\left(
\frac{-2ABD+(A+B)Dr}{8\pi r^5}\right)^{1/n} \right|.
\end{equation}

If $D=10$, $A=1$, and $B=1$, then $r_0=7.5$\,m, while the
condition $b(r)/r=0.001$ produces $a=10000$\,m, and
\[
   n=\frac{37.1}{\text{ln}\,\omega}.
\]

\subsection{$b(r)=exp(E-(Fr^m)^{-1})$}
\begin{equation}
\rho=\frac{m}{8\pi F}\exp[E-(Fr^m)^{-1}]r^{-(m+3)}.
\end{equation}
\begin{equation}
p=-\omega\left[ \frac{m}{8\pi F}\exp[E-(Fr^m)^{-1}]r^{-(m+3)}
\right]^{1+1/n}.
\end{equation}
\begin{equation}
\Phi=\Phi_0-(n+1)\text{ln}\left| 1-\omega\left( \frac{m}{8\pi
F}\exp[E-(Fr^m)^{-1}]r^{-(m+3)}  \right) ^{1/n}\right|.
\end{equation}

If $E=2.1$, $F=1$, and $m=1/2$, then $r_0=5.3$\,m, and the
condition $b(r)/r=0.001$ yields $a=8076$\, m, and
\[
    n=\frac{33.3}{\text{ln}\,\omega}.
\]

\section{Discussion}
This paper generalizes the work of several of our colleagues
\cite{sushkov, fL05, fL06} from the barotropic equation of
state to the polytropic equation $p_r=-\omega\rho^{\gamma}$,
where $\gamma=1+1/n$.  Our analysis results in significant
generalizations of the phantom-energy and, in some cases,
the Chaplygin-gas models, both of which continue
to be studied extensively.  What is of particular interest
is that these generalizations  could have far-reaching
cosmological consequences:  future observations could favor
the new models and thereby increase the chances of finding
naturally occurring wormholes, especially if much of the
Universe is pervaded by such matter.  In the course of the
analysis, several specific solutions were explored: (1) a
constant redshift function, (2) a shape function in the
form of a power, and (3) an isotropic pressure model with
several shape functions.

\vspace{12pt}\noindent
(1) For a constant redshift function, the parameter $n$ must exceed
$-1$.  More precisely, either (a) $\gamma$ is slightly larger
than unity, which is an extension of the phantom-energy model,
or (b)  $-1<\gamma<0$, which is an extension of the generalized
Chaplygin-gas model. The resulting spacetimes are not
asymptotically flat.

\vspace{12pt}\noindent
(2) If the shape function has the form $b(r)=b_0r^m$, $0<m<1$,
an event horizon can be avoided  provided that
\[
   r_0=\left(\frac{\omega}{(8\pi)^{1/n}}m^{(n+1)/n}\right)
       ^{n/2}.
\]
If, in addition, $\omega>(8\pi r^2/b')^{1/n}$, then the WEC is
violated, resulting in a traversable wormhole.  The concomitant
restriction on $n$ can be relaxed if the interior solution is
joined to an exterior Schwarzschild solution.  (See Sec.
\ref{S:poweragain} for details.)

\vspace{12pt}\noindent
(3) For models with isotropic pressure, the parameters $\rho$, $p$,
and $\Phi$ can be readily determined for the shape functions
considered.  If $\omega$ meets the above condition, then the
spacetime is asymptotically flat for any fixed $n$. In most
cases, however, the wormhole cannot be arbitrarily large and
must be cut off and joined to an external Schwarzschild
spacetime.

Various shape functions are discussed, all leading to
$0<\gamma<\gamma_1$, where $\gamma_1>1$.  The magnitude of
$\gamma_1$ depends on the particular shape function.  Most
of these cases are significant extensions of the phantom-energy
model.
\vspace{12pt}

{ \bf Acknowledgements }

We would like to thank the referees for their useful comments to
improve this work. One of us (F.R.) wishes to thank UGC, Government
of India, for providing financial support.

\end{document}